\documentclass[aps,prd,twocolumn,showpacs,superscriptaddress,nofootinbib]{revtex4-1}

\usepackage{amsmath}
\usepackage{amssymb}
\usepackage{graphicx}
\usepackage{hyperref}
\usepackage{bm}
\usepackage{color}

\begin{document}

\title{Why the Brain Cannot Be a Digital Computer:\\
History-Dependence and the Computational Limits of Consciousness}

\author{Andrew F. Knight, J.D.}
\email{aknight@alum.mit.edu}
\date{\today}

\begin{abstract}
This paper presents a novel information-theoretic proof demonstrating
that the human brain as currently understood cannot function as a
classical digital computer. Through systematic quantification of
distinguishable conscious states and their historical dependencies, we
establish that the minimum information required to specify a conscious
state exceeds the physical information capacity of the human brain by a
significant factor. Our analysis calculates the bit-length requirements
for representing consciously distinguishable sensory ``stimulus frames''
and demonstrates that consciousness exhibits mandatory
temporal-historical dependencies that multiply these requirements beyond
the brain's storage capabilities. This mathematical approach offers new
insights into the fundamental limitations of computational models of
consciousness and suggests that non-classical information processing
mechanisms may be necessary to account for conscious experience.
\end{abstract}

\maketitle

\section{Introduction}

The question of whether the human brain operates as a form of digital
computer has remained central to cognitive science, artificial
intelligence research, and philosophy of mind for decades. This
computational theory of mind has influenced neuroscience research
programs, technological development, and philosophical conceptions of
consciousness \cite{piccinini2013}. While various critiques of this
computational paradigm exist \cite{penrose1989,searle1992}, this paper
presents a novel information-theoretic proof demonstrating that the
brain as currently understood cannot function as a classical digital
computer based on a quantitative analysis of conscious states.

A fundamental premise of our argument rests on the principle that
physically distinguishable conscious states must correspond to
physically distinguishable brain states. If two conscious experiences
are subjectively distinct, then (assuming consciousness is produced by
the brain) there must be distinct physical configurations of the brain
corresponding to these experiences. By calculating the minimum
information required to specify distinguishable conscious experiences
across a human lifetime and comparing this to established estimates of
the brain's information capacity, we can determine whether a digital
computational model of the brain is viable.

This approach differs from previous critiques by directly quantifying
the information requirements of consciousness itself rather than
focusing on particular computational limitations or philosophical
arguments. The analysis proceeds from empirically established sensory
discrimination capacities, information compression principles, and
temporal dependencies in conscious experience.

\section{Theoretical Framework}

\subsection{Information Requirements of Conscious States}

To establish the minimum information required to specify a conscious
state, we must first define a ``stimulus frame'' as the minimum unit of
consciously distinguishable sensation across all sensory modalities.
This represents the baseline sensory information that could potentially
be consciously distinguished at a discrete temporal instant. The
conscious experience of a stimulus frame, however, is likely vastly
richer -- a qualitative, integrated phenomenological state that
transcends the mere quantitative specification of sensory data. This
fundamental distinction between sensory input and conscious experience
forms the foundation of our subsequent analysis.

The bit-length ($L$) required to specify a stimulus frame can be
calculated by determining the information requirements for each sensory
modality and summing them. This approach aligns with established methods
in information theory \cite{shannon1948} and builds upon empirical research
in sensory discrimination \cite{clark2013}.

\subsection{Temporal Dependencies in Conscious Experience: The Historical Embeddedness of Consciousness}

A critical aspect of our analysis---and indeed, the fundamental insight
that distinguishes our approach---is the recognition that conscious
states exhibit mandatory historical dependencies. The conscious
experience at time $t_1$ is not determined solely by the sensory
information present at $t_1$, but is contingent on previous experiences at
$t_0$, $t_{-1}$, etc. This historical dependency is not merely an incidental
feature but may actually constitute a fundamental structural property of
consciousness itself \cite{knight2019}.

\subsubsection{The Phenomenology of Sequential Experience}

Referring now to Figure 1, consider experiencing a sunset on a beach:
observing golden light reflecting off ocean waves, hearing the rhythmic
crash of water, tasting wine, smelling salt air, and feeling sand
between your toes. If we compare two scenarios:

\textbf{Scenario 1}: At time $t_0$, you're drinking red wine. At time $t_1$, you
continue drinking the same red wine, producing a coherent experience.

\textbf{Scenario 2}: At time $t_0$, you're drinking white wine. At time $t_1$, you
unexpectedly taste red wine, producing a discontinuous experience.

The sensory input at $t_1$ is identical in both scenarios---same sunset,
same sounds, same red wine. Yet your conscious experience would be
dramatically different. In the first case, you'd likely remain in a
state of peaceful contemplation. In the second, you'd experience
surprise, confusion, or possibly concern that someone switched your
drink.

This thought experiment demonstrates that your conscious state at $t_1$
cannot be specified by the sensory information at $t_1$ alone. It
fundamentally depends on what came before. This phenomenon cannot be
explained away as merely a function of memory or attention. Even without
explicit recollection, the phenomenological character of experience
itself changes based on its historical antecedents. This suggests that
consciousness fundamentally integrates historical information in its
very structure.

As illustrated in Figure 1, this integration exhibits a recursive nested
structure that compounds the information requirements for specifying
conscious states.

\begin{figure*}
\centering
\includegraphics[width=\textwidth]{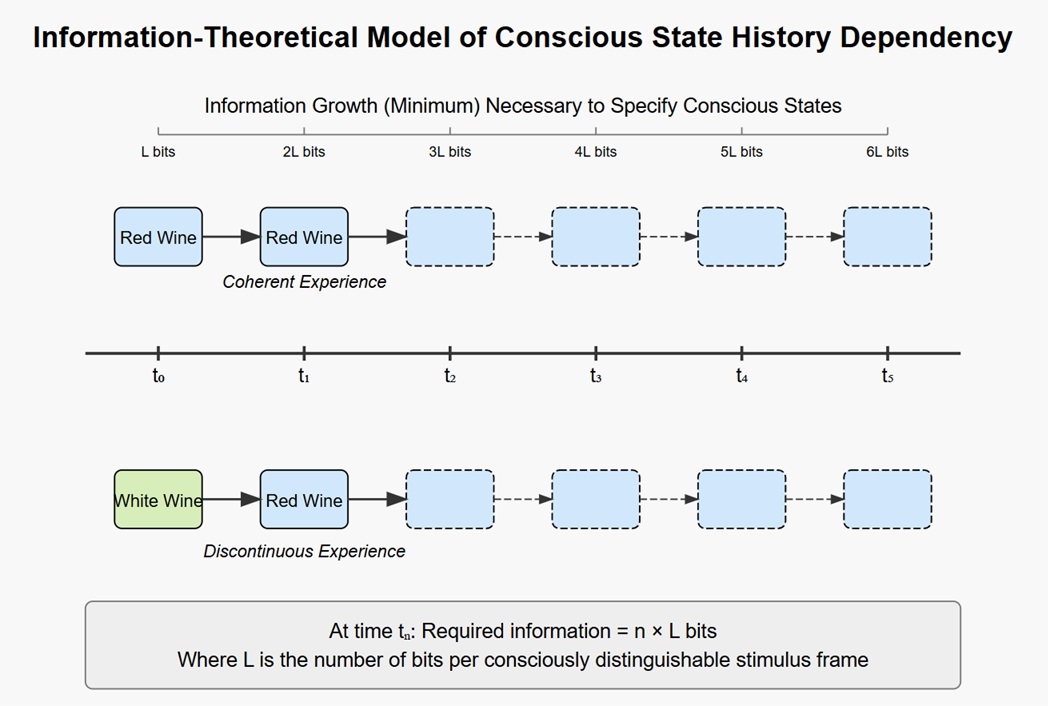}
\caption{Information-Theoretical Model of Conscious State History Dependency.}
\label{fig:model}
\end{figure*}

\subsubsection{Formal Analysis of Historical Embeddedness}

The critical insight emerges when we consider whether a conscious state
at time $t_1$ can be fully determined by the specification of the stimulus
frame at $t_1$ alone. The evidence suggests categorically that it cannot.

Let $C(t_1)$ represent the conscious state at time $t_1$, and $S(t_1)$ represent
the stimulus frame at time $t_1$.

The conventional computational model would suggest: $C(t_1) = f(S(t_1))$

Where $f$ is some function transforming stimulus into consciousness,
requiring $L$ bits to specify.

However, our analysis of the above example demonstrates: $C(t_1) = g(S(t_1), S(t_0))$

Where $g$ is a function integrating both the current and previous stimulus
frames, requiring at minimum $2L$ bits to specify.

This formal representation captures the qualitative difference in
conscious experience between our two scenarios. Despite identical
sensory input $S(t_1)$, the conscious states $C(t_1)$ differ substantially
based on the preceding stimulus frame $S(t_0)$.

\subsubsection{Recursive Integration of Temporal Sequences}

The temporal/historical dependency extends beyond immediate
predecessors, revealing a recursive structure of consciousness. If we
extend our analysis to time $t_2$, the conscious state experienced depends
not merely on the stimulus frame at $t_2$ and its immediate predecessor at
$t_1$, but also on the relationship between stimulus frames and conscious
experiences at $t_1$ and $t_0$.

Consider a sequence where, after experiencing the transition from white
wine to red wine ($t_0 \rightarrow t_1$), the subject continues tasting red wine at $t_2$.
This experience differs qualitatively from one where the subject has
consistently tasted red wine throughout ($t_0 \rightarrow t_1 \rightarrow t_2$). In the former case,
the subject may experience lingering surprise, cognitive reassessment,
or comparison between the wines. This demonstrates that conscious states
at $t_2$ depend on the ordered relationship between previous states.

The history dependence of consciousness follows a recursive pattern:

\begin{enumerate}
\item One's conscious experience at $t_1$ depends on one's conscious
experience at $t_0$.

\item One's conscious experience at $t_3$ depends on one's conscious
experience at $t_2$ (which itself depends on experiences at $t_1$ and $t_0$).

\item The relationship between these two groups matters: one's experience
at $t_3$ would differ if the group of experiences at $t_0$ and $t_1$ came
after the group of experiences at $t_2$ and $t_3$, rather than before.
This relationship thus creates a new layer of dependence.

\item The relationship between these two larger groups matters: one's
experience at $t_7$ would differ if the group of four experiences
through $t_3$ came after the group of four experiences through $t_7$,
rather than before. This relationship creates another layer of
dependence.

\item This pattern continues, with each new grouping creating higher-order
dependencies.
\end{enumerate}

We can illustrate this with a simple mathematical progression:

\begin{itemize}
\item To specify the conscious state at $t_1$ requires (at minimum)
information about stimulus frames at $t_0$ and $t_1$: $2L$ bits

\item To specify the conscious state at $t_3$ requires (at minimum)
information about stimulus frames at $t_0$, $t_1$, $t_2$, and $t_3$, as well as
their ordering: $4L$ bits

\item To specify the conscious state at $t_7$ requires (at minimum)
information about all preceding states ($t_0$ through $t_7$) and their
ordering: $8L$ bits
\end{itemize}

Following this recursive logic, we can formalize the minimum bit length
required to specify a conscious state at time $t_n$ as $n \times L$, where $n$
represents the number of consciously distinguishable stimulus frames
experienced. This formulation reveals the exponential information growth
inherent in conscious experience.

\subsubsection{Mathematical Formalization of Historical Integration}

More generally, the mathematical structure of this historical
integration can be formalized as follows:

Let $C(t_n)$ represent the conscious state at time $t_n$, and $S(t_n)$ represent
the stimulus frame at time $t_n$.

The conventional computational model would suggest: $C(t_n) = f(S(t_n))$

Where $f$ is some function transforming stimulus into consciousness,
requiring $L$ bits to specify.

However, our analysis demonstrates: $C(t_n) = g(S(t_n), S(t_{n-1}), S(t_{n-2}), \ldots, S(t_0))$

Where $g$ is a function integrating the entire historical sequence of
stimulus frames, requiring $n \times L$ bits to specify.

The historical integration is not separable from the immediate
experience---they constitute a unified phenomenological whole.

\subsubsection{Independence from Explicit Memory}

Crucially, this history dependence doesn't require explicit memory. It
makes no difference whether the person actually remembers the particular
stimuli or their order of progression. As long as one has a conscious
experience at, say, time $t_9$ that is in some (even miniscule) manner
dependent on the preceding nine stimulus frames and their order, then it
takes at least ten sets of bit strings of length $L$ to specify.

This can be understood through analogy to cinematic experience. The
emotional impact and comprehension of the final scene of a film depend
on all preceding scenes, even if viewers cannot explicitly recall each
scene. The phenomenological experience of watching the conclusion
incorporates the entire narrative history in a manner that cannot be
reduced to the visual and auditory information of the final scene, or
even the last few scenes, alone.

The specification of each conscious experiential moment requires not
just the current sensory frame but the entire ordered sequence of
previous frames and their relationships. This recursive embedding means
conscious states have an inherently nested temporal/historical
structure.

\subsubsection{Implications for Information Theory of Consciousness}

This historical embeddedness of consciousness has profound implications
for any information-theoretic analysis of mental states. If conscious
states necessarily incorporate their historical antecedents, then the
minimum information required to specify any given conscious state must
include information about all preceding states.

This stands in stark contrast to traditional computational models of
cognition that treat mental states as functions of current input plus
some fixed set of parameters. The recursive temporal integration we have
identified suggests that conscious states have an intrinsically
historical structure that cannot be reduced to or compressed into
ahistorical computational processes.

As we will demonstrate in subsequent sections, this historical
dependency creates information requirements that exceed the physical
storage capacity of the brain when modeled as a classical digital
computer. This poses a fundamental challenge to computational theories
of mind and necessitates either a radical revision of our understanding
of neural information processing or the recognition that consciousness
cannot be fully explained within a classical computational framework.

\section{Quantifying Sensory Information Content}

To rigorously establish the information content required to specify
distinguishable conscious states, we must systematically analyze each
sensory modality. This quantification provides the foundation for our
subsequent analysis of the brain's information capacity limits. Our
methodological approach integrates empirical psychophysical data with
information-theoretic principles to establish conservative lower bounds
for sensory information requirements.

\subsection{Visual Information}

The human visual system represents the most information-intensive
sensory modality in terms of neural resources and processing capacity
\cite{vanessen1995}. To establish the minimum information
required to specify a distinguishable visual frame, we begin with the
standard high-definition visual resolution.

A 1080p frame comprises $1920 \times 1080$ pixels, with each pixel encoding
three color channels (red, green, blue) at 8 bits per channel. This
yields a raw information content of 49,766,400 bits.

This calculation represents the uncompressed information content of a
single visual frame. However, this raw calculation significantly
overestimates the information required to specify a consciously
distinguishable visual frame, as human visual perception exhibits
numerous limitations and efficiencies.

Digital video compression algorithms such as H.264/AVC and H.265/HEVC
achieve compression ratios of 300:1 or greater while maintaining
perceptual quality that appears visually indistinguishable to human
observers \cite{wiegand2003,sullivan2012}. These
algorithms exploit multiple properties of human visual perception, such
as spatial redundancy (whereby adjacent pixels in natural scenes tend to
be highly correlated, allowing for efficient encoding of spatial
patterns \cite{balle2018}) and color perception constraints (though
humans can theoretically distinguish millions of colors, the actual
discriminability under natural viewing conditions is substantially lower
\cite{fairchild2013}).

Compression algorithms utilizing these principles can reduce a 1080p
frame at 30fps to approximately 8 Mbps (megabits per second) without
perceptible degradation to most observers. This corresponds to:

$8,000,000 \text{ bits/second} \div 30 \text{ frames/second} \approx 267,000 \text{ bits per frame}$

However, this still likely overestimates the minimum requirements for a
consciously distinguishable visual frame. Psychophysical research by
Pitkow and Meister \cite{pitkow2014} suggests that natural scene statistics allow
for even greater compression when optimized specifically for human
perceptual limitations rather than technological constraints.

Taking a conservative approach and further reducing this estimate by a
factor of 10 to account for additional redundancies and perceptual
limitations yields approximately 25,000 bits per consciously
distinguishable visual frame. This estimate aligns with
information-theoretic analyses of visual perception by Field \cite{field1987} and
more recent work by Moreno-Bote et al. \cite{morenobote2014}, who demonstrated that
visual perception operates at approximately 5-10\% of the theoretical
channel capacity limit of the retinal output.

The range of plausible values based on current research spans from
approximately 10,000 to 50,000 bits per consciously distinguishable
visual frame, with our analysis adopting 25,000 bits as a well-supported
midpoint estimate.

\subsection{Auditory Information}

Unlike visual information, auditory perception is fundamentally
temporal, requiring analysis of pressure variations over time. Standard
digital audio employs sampling rates of 44.1-48 kHz with 16-24 bit depth
per sample \cite{brandenburg1994}.

For a sampling rate of 48 kHz, each second contains 48,000 samples.
Considering the established psychophysical finding that humans can
distinguish visual frames at approximately 30 frames per second
\cite{holcombe2009}, we can correspondingly partition auditory information
into equivalent temporal units:

$48,000 \text{ samples/second} \div 30 \text{ frames/second} \approx 1,600 \text{ samples per ``auditory frame''}$

With 16-bit encoding, this yields: $1,600 \text{ samples} \times 16 \text{ bits} = 25,600 \text{ bits per raw auditory frame}$

However, as with visual information, this raw calculation substantially
overestimates the information required to specify a consciously
distinguishable auditory experience. Perceptual audio coding algorithms
like MP3 and AAC achieve compression ratios of 10:1 or greater while
maintaining perceptual transparency for most listeners \cite{brandenburg1999}. These
algorithms exploit a variety of psychoacoustic principles, such as
frequency masking, whereby louder sounds mask quieter sounds at
similar frequencies \cite{moore2012}.

MP3 encoding at 96 kbps (kilobits per second) yields: $96,000 \text{ bits/second} \div 30 \text{ frames/second} \approx 3,200 \text{ bits per auditory frame}$

Following the same conservative approach used for visual information and
reducing by an additional factor of 10 to account for further perceptual
limitations, we estimate approximately 320 bits per consciously
distinguishable auditory frame. This estimate aligns with research by
McDermott and Simoncelli \cite{mcdermott2011}, who demonstrated that auditory
perception of natural sounds relies on a relatively small set of
statistical parameters.

\subsection{Olfactory Information}

Olfactory perception has traditionally been considered limited in
information capacity compared to vision and audition. However, recent
research has substantially revised this understanding.

Bushdid et al. \cite{bushdid2014} conducted a landmark study demonstrating that
humans can discriminate at least 1 trillion distinct odors, orders of
magnitude greater than previous estimates. This discrimination capacity
can be represented information-theoretically as:

$2^{40} \approx 1 \text{ trillion distinct odors}$

Therefore, approximately 40 bits of information are required to uniquely
specify any distinguishable odor within human perceptual capacity. While
some researchers have suggested that the effective dimensionality of
human olfactory perception may be lower \cite{koulakov2011}, the
40-bit estimate represents a conservative lower bound based on empirical
discrimination data.

\subsection{Gustatory Information}

Taste perception involves the discrimination of compounds across five
primary taste qualities (sweet, sour, salty, bitter, umami) and their
combinations. Research by Peng et al. \cite{peng2015} indicates that humans can
discriminate approximately 100,000 to 1,000,000 distinct taste
experiences when considering combinations of primary tastes and their
intensities. This corresponds to approximately:

$2^{20} \approx 1 \text{ million distinct taste experiences}$

Therefore, approximately 20 bits of information are required to specify
a distinguishable gustatory experience. However, considering the complex
integration with olfaction and the findings of Bushdid et al. \cite{bushdid2014}
regarding flavor perception, we adopt a conservative estimate of 40 bits
per distinguishable gustatory frame, equivalent to our olfactory
estimate.

\subsection{Tactile Information}

Tactile perception encompasses pressure, temperature, pain, and pleasure
sensations distributed across the body surface. Quantifying tactile
information requires considering both spatial distribution and
qualitative dimensions of touch.

The human body has a surface area of approximately 1.5-2 m² \cite{montagna1972}. The
minimum distinguishable spatial separation between two
tactile stimuli (two-point discrimination threshold) varies across body
regions, from approximately 1 mm on the fingertips to several
centimeters on the back \cite{weinstein1968}.

Taking a conservative average discrimination threshold of 1 cm across
the body surface, we can calculate approximately 15,000 distinguishable
tactile ``pixels'' across the body surface. Each of these locations can
register various sensory qualities (pressure intensity, temperature,
pain, pleasure) with multiple distinguishable gradations.

Research by Gescheider et al. \cite{gescheider2009} indicates that humans can
distinguish approximately 5-7 levels of pressure intensity under
experimental conditions, while thermal sensitivity studies by Stevens
\cite{stevens1957} demonstrate discrimination of approximately 4-5 distinct
temperature levels within the non-painful range.

Pain perception exhibits greater complexity, with at least 3
distinguishable dimensions (intensity, quality, and temporal
characteristics) according to Price et al. \cite{price1987}. Each dimension
contains multiple distinguishable levels, yielding dozens of
discriminable pain states.

Conservatively estimating 10 bits ($2^{10} = 1,024$ distinct states) per
tactile location to encode all these qualitative dimensions, we
calculate:

$15,000 \text{ tactile pixels} \times 10 \text{ bits/pixel} = 150,000 \text{ bits}$

This estimate aligns with research on somatosensory coding by Bensmaia
and Hollins \cite{bensmaia2005}, who demonstrated that tactile perception involves
complex spatiotemporal patterns of neural activity across multiple
mechanoreceptor types.

More recent work by Weber et al. \cite{weber2013} has further characterized the
information capacity of the human somatosensory system, confirming that
tactile information processing involves parallel channels with distinct
temporal and spatial filtering properties.

\subsection{Integrated Sensory Information}

Summing the information content across all sensory modalities (while
acknowledging the dominance of visual and tactile information), we
arrive at a conservative estimate of approximately 200,000 bits per
integrated stimulus frame. This represents the bit length ($L$) required
to specify the minimum unit of consciously distinguishable sensation
across all modalities.

It is important to note that this estimate:

\begin{enumerate}
\item Is deliberately conservative, representing a lower bound rather than
an upper limit

\item Is based on psychophysical discrimination thresholds rather than
theoretical information capacity

\item Considers only the minimal information required for conscious
discrimination, not the full richness of phenomenological experience

\item Is supported by converging evidence from multiple methodological
approaches
\end{enumerate}

This 200,000-bit estimate provides the foundation for our subsequent
analysis of temporal/historical dependencies and their implications for
the brain's information processing capacity.

\section{Discussion}

\subsection{Total Information Requirements for Conscious Experience}

To calculate the minimum total information requirement for conscious
experience over a human lifetime, we multiplied the bit-length of a
stimulus frame ($L \approx 200,000$ bits) by the total number of distinguishable
frames experienced.

Research on the temporal resolution of human perception suggests that we
can distinguish sensory events separated by approximately 50-70
milliseconds \cite{vanrullen2003,wittmann2011}. This corresponds
to a theoretical maximum of about 15-20 distinguishable frames per
second. Hecht and Verrijp \cite{hecht1933} established that humans can detect
visual flicker up to approximately 15-20 Hz under optimal conditions,
while more recent work by Battelli et al. \cite{battelli2007} corroborates that
temporal individuation of visual events occurs with a resolution of
approximately 50-60 ms.

Taking a conservative estimate of 15 distinguishable frames per second
over a 100-year lifespan:

\begin{itemize}
\item $15 \text{ frames/second} \times 60 \text{ seconds/minute} \times 60 \text{ minutes/hour} \times 24 \text{ hours/day} \times 365.25 \text{ days/year} \times 100 \text{ years} \approx 47.3 \text{ billion frames}$

\item $47.3 \text{ billion frames} \times 200,000 \text{ bits/frame} \approx 9.46 \times 10^{15} \text{ bits (9.46 quadrillion bits)}$
\end{itemize}

As depicted in Figure 1, the recursive temporal integration structure of
consciousness means that each of these frames must be considered in
relation to previous frames, creating the substantial information
requirements calculated here.

\subsection{Comparison with Neural Information Capacity}

To establish a rigorous comparison between the information requirements
of consciousness and the physical constraints of neural architecture, we
must quantify the brain's theoretical information storage capacity.
This necessitates a systematic examination of neuroanatomical parameters
and their information-theoretic implications.

\subsubsection{Neuroanatomical Parameters}

\textbf{Neuronal Population}: The consensus estimate of human brain neuronal
population has undergone significant refinement through methodological
innovations. While earlier estimates varied considerably, contemporary
quantitative analyses employing isotropic fractionation techniques have
yielded more precise values. Herculano-Houzel \cite{herculano2009} established that
the human cerebral cortex contains approximately 16 billion neurons,
with the cerebellum containing an additional 69 billion, yielding a
total of approximately 86 billion neurons. More recent work by von
Bartheld et al. \cite{vonbartheld2016} synthesizing multiple methodologies corroborates
this range, suggesting 80-100 billion neurons as the most reliable
estimate for the complete human brain.

\textbf{Synaptic Density}: The average synaptic connectivity per neuron
exhibits substantial regional and cell-type variation. DeFelipe \cite{defelipe2011}
demonstrated that cortical pyramidal neurons establish approximately
7,000 synaptic connections, while other neuronal types exhibit different
connectivity patterns. Comprehensive analyses by Tang et al. \cite{tang2001} and
Kasthuri et al. \cite{kasthuri2015} utilizing electron microscopy and computational
reconstruction have established a range of 1,000-10,000 synapses per
neuron, with 7,000 representing a neurobiologically plausible mean
value.

\textbf{Synaptic Information Content}: The information content encodable
within a single synapse represents a critical parameter for neural
information capacity calculations. Bartol et al. \cite{bartol2015} demonstrated
that synaptic strength distributions exhibit multimodal characteristics
with approximately 26 distinguishable states (equivalent to 4.7 bits)
per hippocampal synapse. This aligns with theoretical work by Varshney
et al. \cite{varshney2006}, which established information-theoretic bounds on
synaptic efficacy coding at approximately 4.5-5 bits.

\subsubsection{Theoretical Information Capacity}

Given these refined parameters, we can calculate the theoretical upper
bound on neural information storage capacity through two complementary
approaches:

\begin{enumerate}
\item \textbf{Neurons as Binary Units}: Under the simplifying assumption that
neurons function as binary units (active/inactive), the maximum
theoretical information capacity would be:

$8.6 \times 10^{10} \text{ neurons} \times 1 \text{ bit/neuron} \approx 8.6 \times 10^{10} \text{ bits (86 billion bits)}$

This capacity falls short of our calculated requirement ($9.46 \times 10^{15}$
bits) by over five orders of magnitude.

\item \textbf{Synaptic Information Storage}: A more neurobiologically plausible
model incorporates synaptic weights as the primary information
substrate:

$8.6 \times 10^{10} \text{ neurons} \times 7,000 \text{ synapses/neuron} \times 4.7 \text{ bits/synapse} \approx 2.8 \times 10^{15} \text{ bits (2.8 quadrillion bits)}$

While substantially greater than the neuronal binary model, this
capacity remains inadequate relative to our (very conservatively)
calculated requirement by a factor of approximately 3.4.
\end{enumerate}

These calculations align with independent analyses by Landauer \cite{landauer2001},
who estimated the brain's lifetime information storage capacity at
approximately $10^{13}$-$10^{14}$ bits through psychophysical measures of human
learning and memory. More recent theoretical work by Bartol et al.
\cite{bartol2015} suggests an upper bound of approximately $2.7 \times 10^{15}$ bits when
accounting for structural plasticity and multisynaptic connections.

Anatomically detailed simulations by Markram et al. \cite{markram2015} as part of
the Blue Brain Project yielded compatible estimates, suggesting that
even when incorporating dendritic computational properties and
sub-synaptic molecular information storage mechanisms, the theoretical
capacity remains below $5 \times 10^{15}$ bits.

\subsubsection{Information-Theoretic Implications}

This quantitative discrepancy between required information capacity
($9.46 \times 10^{15}$ bits) and the brain's theoretical storage capacity ($2.8 \times 10^{15}$ bits) represents a factor of 3.4---a significant disparity when
considering the deliberately conservative approach employed in our
analysis.

Several critical observations emerge from this comparison:

\begin{enumerate}
\item Even adopting optimistic estimates of neural information capacity,
the brain appears fundamentally inadequate as a classical digital
information storage medium capable of encoding the complete set of
distinguishable conscious states experienced over a human lifetime.

\item The magnitude of this discrepancy suggests that the limitation is
not merely a matter of parametric refinement but represents a
fundamental constraint on classical neural information processing.

\item The disparity remains robust even when incorporating
state-of-the-art understanding of neuronal and synaptic information
encoding mechanisms.
\end{enumerate}

This analysis provides quantitative support for our central thesis that
the brain as currently understood cannot function as a classical digital
computer with respect to consciousness. The minimum information
requirements for specifying conscious states throughout a human lifetime
exceed the physical constraints of neural architecture by a significant
margin, necessitating alternative theoretical frameworks for
understanding the neural basis of consciousness.

\section{Implications}

\subsection{Philosophical Implications of Historical Embeddedness}

The temporal-historical dependency of conscious states elucidated in our
analysis carries profound philosophical implications that extend beyond
computational limitations, challenging fundamental assumptions about the
nature of consciousness itself.

\subsubsection{Ontological Status of Conscious States}

The historical embeddedness of consciousness necessitates a radical
reevaluation of the ontological status of conscious states. Conventional
philosophical frameworks---including functionalist, representationalist,
and certain physicalist theories---implicitly presuppose that conscious
states can be individuated as discrete temporal entities. Our analysis
demonstrates that this presupposition is fundamentally untenable.

If consciousness necessarily integrates its entire historical sequence,
then it cannot be adequately characterized as a momentary state or as a
sequence of independent states. Rather, consciousness exhibits what
might be termed ``temporal holism''---a structural property wherein each
experiential moment necessarily incorporates the unified temporal
gestalt of prior experience. This challenges the metaphysical adequacy
of atomistic approaches to conscious states and suggests that
consciousness possesses an inherently holistic temporal structure that
resists decomposition into temporally discrete units.

This temporal holism bears significant parallels to the holistic
features identified in quantum mechanical systems, where the state of a
composite system cannot be reduced to the states of its constituent
components. Just as quantum entanglement necessitates describing certain
physical systems as irreducible (non-separable) wholes, the historical
embeddedness of consciousness may require treating the temporal
extension of consciousness as an irreducible unity.

\subsubsection{Challenge to Stimulus-Response Models}

The history dependence argument fundamentally challenges simplistic
stimulus-response models of consciousness. Our analysis demonstrates
mathematically that consciousness cannot be reduced to a function
mapping current stimuli to current states. The formal representation:

$C(t_n) = g(S(t_n), S(t_{n-1}), S(t_{n-2}), \ldots, S(t_0))$

reveals that any adequate theory of consciousness must account for the
integration of historical information into present experience. This
challenges not only computational theories of mind but also various
forms of representationalism that characterize conscious states
primarily in terms of their representational relationship to current
environmental conditions.

The inadequacy of stimulus-response models has significant implications
for cognitive science and artificial intelligence research. Contemporary
neural network architectures---including recurrent networks with memory
mechanisms---remain fundamentally limited in their capacity to integrate
historical information compared to the recursive temporal embedding
exhibited by consciousness.

\subsubsection{Experiential Richness Through Historical Integration}

The historical dependency structure of consciousness provides a rigorous
explanation for the phenomenological richness and depth of conscious
experience. Our conscious states possess qualitative depth and nuance
precisely because they integrate vast histories of experience. This
integration is not merely additive but involves complex recursive
relationships between experiential elements across multiple temporal
scales.

This account offers a formal framework for understanding phenomena such
as experiential depth, aesthetic appreciation, and emotional resonance.
The recursive temporal structure we have identified provides a
mathematical and logical basis for these phenomenological observations.
Each conscious moment recursively incorporates prior conscious states,
creating a nested temporal structure of extraordinary complexity. This
nested structure allows for the emergence of higher-order experiential
properties that cannot be reduced to or derived from individual stimulus
frames in isolation.

\subsection{Implications for Computational Models of Consciousness}

Our findings present significant challenges to classical computational
theories of mind. If consciousness requires more information than the
brain can physically encode, then either:

\begin{enumerate}
\item Consciousness is not entirely produced by the brain;

\item The brain utilizes non-classical computational mechanisms that
transcend traditional information theory constraints; or

\item Current estimates of neural information capacity are fundamentally
inadequate.
\end{enumerate}

The first possibility aligns with various dualist or non-physicalist
theories of consciousness \cite{chalmers1996}. The second possibility
suggests quantum computational models \cite{hameroff2014} or
other non-classical information processing mechanisms might be
necessary. The third possibility would require radical revisions to our
understanding of neural information encoding.

\section{Conclusion}

This paper presents a novel information-theoretic proof demonstrating
that the human brain as currently understood cannot function as a
classical digital computer. By quantifying the minimum information
required to specify distinguishable conscious states over a human
lifetime and comparing this to the brain's physical information
capacity, we have shown that there exists a fundamental discrepancy that
challenges computational theories of mind.

This analysis does not deny that the brain performs information
processing or computation broadly defined. Rather, it specifically
demonstrates that classical digital computational models are
insufficient to account for the full range of distinguishable conscious
experiences. This suggests that either consciousness extends beyond the
brain, the brain utilizes non-classical computational mechanisms, or our
understanding of neural information encoding requires radical revision.

These findings contribute to ongoing debates in philosophy of mind,
cognitive science, and artificial intelligence, offering quantitative
constraints that any viable theory of consciousness must satisfy. Future
research should focus on developing and testing alternative models of
consciousness that can operate within the physical constraints of neural
systems while accounting for the history dependence of consciousness
that gives rise to the rich tapestry of human experience.

\end{document}